\documentclass[sigconf,natbib=false]{acmart}

\setcopyright{acmlicensed}
\usepackage{amsmath,amsfonts}
\usepackage{algorithm}
\usepackage{algorithmic}
\usepackage{graphicx}
\usepackage{textcomp}
\usepackage{xcolor}
\usepackage{subcaption}
\usepackage{tabularx}
\usepackage{ragged2e}
\usepackage{pdfpages}
\usepackage{float}
\usepackage{comment}
\usepackage{tikz}
\usetikzlibrary{arrows.meta,calc,positioning,fit,decorations.pathreplacing,decorations.pathmorphing}
\usepackage{enumitem}
\usepackage{longtable}
\usepackage{booktabs}
\usepackage{placeins}
\usepackage{soul}
\usepackage{bbold}
\usepackage{siunitx}
\DeclareSIUnit\molar{M}

\sethlcolor{yellow}
\renewcommand{\hl}[1]{#1}

\RequirePackage[
  datamodel=acmdatamodel,
  style=acmnumeric,
  maxbibnames=3,
  minbibnames=1,
  maxcitenames=2
]{biblatex}

\title[Whole Blood BioFET ctDNA Boundary Analysis of BioFET-Based ctDNA Detection]{Whole-Blood Boundary Analysis of BioFET-Based ctDNA Detection for Intravascular Sensing in Intrabody Nanonetworks}

\author{Ida Kleger-Rudomin}

\affiliation{%
  \institution{Gdansk University of Technology}
  \country{Poland}
}

\author{Filip Lemic}

\affiliation{%
  \institution{i2Cat Foundation}
  \country{Spain}
}
\author{Sergi Abadal}
\affiliation{%
  \institution{Universitat Polit\`ecnica de Catalunya}
  \country{Spain}
}
\author{Eduard Alarc\'on}
\affiliation{%
  \institution{Universitat Polit\`ecnica de Catalunya}
  \country{Spain}
}

\author{Ethungshan Shitiri}
\affiliation{%
  \institution{Universitat Polit\`ecnica de Catalunya}
  \country{Spain}
}
\authornote{Corresponding author}

\begin{document}

\begin{abstract}
Liquid biopsy can detect tumor-derived biomarkers such as circulating tumor DNA (ctDNA), but ultra-low-fraction assays remain costly, slow, and difficult to scale. This motivates interest in intravascular in vivo sensing in the context of intrabody nanonetworks, where nanosensors could support local biomarker monitoring. BioFET-based nanosensors are relevant here because they are label-free, highly miniaturizable, and have shown strong ctDNA sensitivity in controlled media. We examine whether this sensitivity still yields reliable ctDNA detection in whole blood using a reduced-order stochastic simulation model that links operating-point selection, Debye-screened charge transduction, stochastic finite-capacity binding, nonspecific adsorption, background fluctuations, and intrinsic electronic noise to blank-threshold detection. Monte Carlo evaluation under the stated physiological screening and lumped interface assumptions shows that short Debye length and interface attenuation reduce the current shift, while low-frequency noise and background fluctuations reduce the margin between target-present and blank responses.
Across the tested quasi-static charge-gating regimes, sensitivity ranges from low values under short Debye length and thicker interfaces to near 100\% for higher target concentrations under stronger coupling conditions. Specificity ranges from approximately 85\% to 91\% across the evaluated oxide thicknesses. The model therefore provides an exploratory whole-blood boundary analysis that identifies how screening, interface geometry, target concentration, and electronic noise shape BioFET-based intravascular ctDNA detection.
\end{abstract}



\copyrightyear{2026}
\acmYear{2026}
\setcopyright{cc}
\setcctype{by}
\acmConference[NANOCOM '26]{13th Annual ACM International Conference on Nanoscale Computing and Communication}{September 21--23, 2026}{St. John's, NL, Canada}
\acmBooktitle{13th Annual ACM International Conference on Nanoscale Computing and Communication (NANOCOM '26), September 21--23, 2026, St. John's, NL, Canada}
\acmDOI{10.1145/3818305.3830255}
\acmISBN{979-8-4007-2767-2/2026/09}

\maketitle

\section{Introduction}
Early-stage cancer detection is important in oncology because outcomes improve when cancers are identified earlier, yet many are still diagnosed at advanced stages \cite{Liang2025LB}. This need has increased interest in blood-based molecular diagnostics that aim to detect rare tumor-derived alterations in noisy clinical backgrounds. Liquid biopsy has therefore emerged as an important route to diagnosis and monitoring by enabling access to tumor-derived material from blood without invasive tissue procedures. Among liquid-biopsy analytes, circulating tumor DNA (ctDNA) is especially informative because it carries cancer-specific genetic alterations and supports quantitative tracking of disease dynamics. The most demanding applications require detecting ctDNA at very low fractions amid a large background of non-tumor cell-free DNA (cfDNA), especially in early-stage disease and after treatment \cite{Ma2024LB_Rev,siravegna2017integrating}. In this regime, performance depends on mean signal level, counting uncertainty, and background interference. These factors increase assay burden and motivate sensing strategies that can reduce reliance on intermittent laboratory measurements.

\begin{figure*}[!t]
    \centering
    \includegraphics[width=0.80\textwidth]{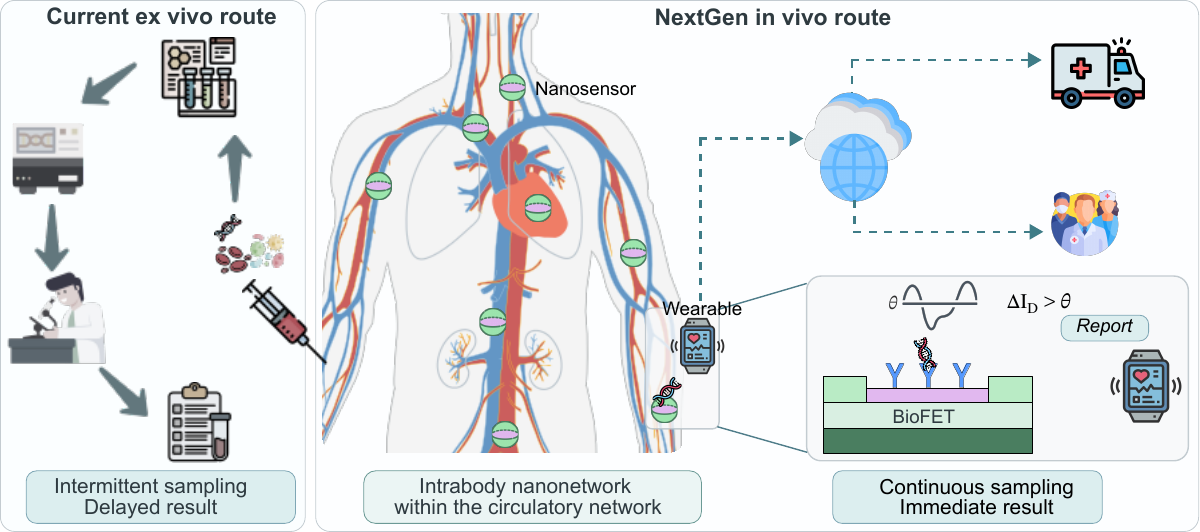}
    \caption{Whole-blood ctDNA sensing in an intrabody nanonetwork compared with conventional liquid biopsy.}\vspace{-1em}
    \label{fig_intro_bottleneck}
\end{figure*}

\hl{One response to this burden is to move sensing closer to the physiological site of interest for \textit{in situ} molecular detection. Intravascular sensing can reduce the delay between biomarker release and measurement, avoid ex vivo sample collection and pre-analytical transport artifacts, and allow circulating nanosensors to sample regions near the tumor origin, where biomarker concentrations may be locally elevated} \cite{Shitiri2024IBN,Frutiger2021NSB,siravegna2017integrating}. In this context, intrabody nanonetworks provide the broader systems context for this direction. In that framework, nanosensors detect local biochemical events and produce signals that can, in principle, be digitized and reported under tight size, power, and duty-cycling constraints \cite{Akyildiz2015IOBNT}. 
\hl{Biological field-effect transistor (BioFET) architectures are established in molecular communication (MC) as high-performance receiver models. Prior work offers analytical end-to-end frequency-domain models for microfluidic channels} \cite{Abdali2024} \hl{and microfluidic experiments with graphene receivers and ssDNA-encoded messages} \cite{Kuscu2021}. \hl{Recent MC receiver design advances include frequency-domain detection using binding noise power spectral density to distinguish signal from interference} \cite{Civas2024}, alongside alternative transduction mechanisms addressing Debye screening and reference electrode requirements \cite{Aktas2021}.

\hl{However, existing studies often bypass sample complexity using simplified media or ex vivo purification, neglecting the fundamental Debye screening and nonspecific fouling barriers that hinder direct intravascular operation} \cite{li2021supersensitive, Stern2010WholeBlood, Kesler2020DebyeBeyond}. \hl{This work extends existing receiver frameworks by providing a whole-blood boundary analysis for intravascular sensing, incorporating the combined impact of short Debye length, competitive surface occupancy, and device noise on binary detection limits for ultra-low ctDNA concentrations. Specifically, this study evaluates the separation between nanosensor responses with ctDNA present and target-absent whole-blood responses under physiological screening, background interference, and device noise.} Fig.~\ref{fig_intro_bottleneck} summarizes this intrabody setting and the bottlenecks that motivate the exploration undertaken here.

BioFETs are transistor-based nanosensors in which biomolecular binding at a functionalized surface changes the local electrostatic potential and shifts the channel current \cite{Lowe2017FieldEffect}. In the intravascular sensing setting, BioFETs are relevant because they are label-free, electrically integrated, and have already shown strong ctDNA sensitivity in buffer and serum \cite{li2021supersensitive,peng2025signal}. Their direct operation in whole blood remains strongly constrained because short electrolyte Debye length, biofunctional layer separation, oxide mediated gate coupling, nonspecific adsorption, and intrinsic low-frequency noise can interfere with the signal \cite{Stern2010WholeBlood,Kesler2020DebyeBeyond,Frutiger2021NSB,Gasparyan2021InternalNoises}.

Work on high-ionic-strength and complex media has used local screening control, alternative measurement schemes, and interface engineering to recover charge coupling or measurement stability \cite{Gao2015HighIonicStrength,Chu2017BeyondDebyeSerum}. Recent studies have moved toward whole-blood BioFET operation through upstream matrix handling or analyte-specific architectures \cite{Stern2010WholeBlood,Babbar2026WholeBlood,Samanta2026LDopaWholeBlood}. The broader whole-blood sensing literature also shows that intravascular operation is possible for selected analytes and mechanisms. Electrochemical aptamer-based sensors avoid Debye screening and have enabled seconds-resolved in vivo molecular measurements in circulating blood \cite{downs2022realtime}. Enzyme amperometric sensors support long-duration metabolic monitoring in vivo \cite{wu2023deviceintegration}. Fiber-optic photoacoustic sensors have demonstrated intravascular whole-blood sensing for analytes such as dissolved gases and heparin \cite{zhou2022heparin}. We therefore use BioFET as a charge-based reference platform whose controlled-media ctDNA sensitivity and electrical compatibility make it informative for a whole-blood boundary analysis, while recognizing that direct whole-blood ctDNA sensing remains unestablished across platforms.

We address this transfer question with a reduced-order stochastic simulation model for BioFET-based ctDNA sensing under practical whole-blood conditions. The model links operating-point selection, interface geometry, finite-capacity binding, screened transduction, background response, noise, and threshold-based detection in a single local-sensing pipeline. The main contributions are as follows.
\begin{itemize}[leftmargin=*,topsep=2pt,itemsep=1pt,parsep=0pt]
    \item We define intravascular ctDNA sensing as a blank-threshold detection problem under intrabody-compatible local sensing constraints, where detection depends on separation between target-present responses and target-absent whole-blood blank responses.
    \item We develop a reduced-order stochastic simulation model that combines analytical BioFET charge coupling and blank-threshold detection with Monte Carlo sampling of target binding, cfDNA-like background fragments, and intrinsic current noise.
    \item We quantify how oxide thickness, effective binding distance, ionic screening, and target concentration shape whole-blood detection performance, including regimes in which specificity remains above 85\% and sensitivity changes markedly with concentration and interface conditions.
\end{itemize}

The remainder of the paper is organized as follows. Section~\ref{sec_Framework} presents the stochastic end-to-end system model linking ctDNA concentration to drain-current shifts and binary detection outcomes. Section~\ref{sec_SimResults}  reports simulation results on screening-limited signal formation, sensitivity and specificity across interface and screening regimes, and the low-frequency noise floor that compresses the margin between target-present and blank responses. Section~\ref{sec_conclusion} then discusses the implications of these results, the model limitations, and design directions for future work.

\section{ctDNA Sensing Model}\label{sec_Framework}

\begin{figure}[!t]
    \centering
\includegraphics[width=0.990\columnwidth]{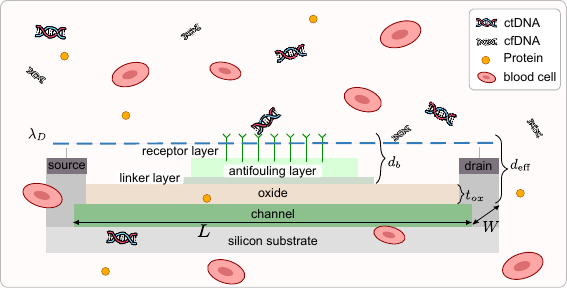}
        \caption{\hl{Modeled BioFET and whole-blood background.}}\vspace{-1em}
 \label{fig_biofet} 
\end{figure}

We define a reduced-order, quasi-static stochastic simulation model that maps ctDNA concentration in whole blood to a drain-current shift and a binary detection outcome. The model focuses on the local sensing stage. Device-specific implementation details and the intrabody reporting chain are outside the present scope. Fig.~\ref{fig_biofet} highlights the modeled sensing stack and the whole-blood environment considered here. The model has four stages. First, it sets the operating point and baseline current. Second, it models surface occupancy by target and background fragments. Third, it converts bound charge into a Debye-screened current shift. Fourth, it estimates a blank-derived threshold and computes sensitivity and specificity. The term \textit{blank} refers to measurements performed in the absence of target molecules, so the response reflects only background and noise contributions.

\subsection{Operating Point and Baseline Current}

We first define the electrical operating point because the sensing model measures current deviations from that baseline. The BioFET is evaluated at $V_{SG}=0.3$ V and $V_{SD}=0.1$ V. \hl{These biases correspond to the assumed fixed subthreshold operating point, where small surface-potential variations produce large relative changes in drain current} \cite{Lowe2017FieldEffect}. In this regime, surface-potential variations are converted into drain-current shifts through the transconductance $g_m$. \hl{The transconductance $g_m$ is evaluated based on the chosen operating point and the given $t_ox$.} Let $I_{D,0}$ denote the baseline drain current at $(V_{SG},V_{SD})$ in the absence of sensing effects.
The measured drain current is therefore written as 
\begin{equation}
I_D = I_{D,0} + \Delta I_D + \eta_I ,
\label{eqID}
\end{equation}
where $I_D$ is the measured drain current, $\Delta I_D$ is the current shift, and $\eta_I$ is the intrinsic current-noise term, defined in Section~\ref{sec_blank_thres}.

\subsection{Sensing Interface and Finite-Capacity Occupancy}

We next model the biofunctionalized surface. In whole blood, ctDNA recognition occurs alongside cfDNA, proteins, and other background molecules. The BioFET therefore responds to both specific ctDNA binding and nonspecific adsorption, meaning unwanted binding by non-target molecules. Passivation can reduce nonspecific adsorption and delay biofouling, which is the gradual formation of an adsorption layer that blocks binding sites, shifts surface charge, and introduces time-dependent drift \cite{Frutiger2021NSB}. For nucleic-acid targets such as ctDNA, selectivity is typically provided by aptamers or ssDNA probes designed for mutation-level discrimination \cite{Wen2022ctDNA,li2021supersensitive}.

In practice, the interface forms a multilayer stack that includes a linker or anchoring chemistry, an antifouling or passivation layer, and the bioreceptor itself \cite{Movilli2018Control}. This stack creates the main interface tradeoff in the model. Passivation suppresses background adsorption at the cost of increasing the distance between target charge and the transducer. Under physiological screening, that added distance weakens electrostatic coupling \cite{Kesler2020DebyeBeyond}. Thinner and less passivated interfaces can increase coupling, although they are more exposed to nonspecific adsorption and drift. For charge-based sensing, even a thin adsorption layer can shift the apparent baseline because the transistor responds to the net electrostatic boundary condition set by both specific target binding and background adsorption.

The present model captures these effects through an abstraction based on competitive surface occupancy, background charge fluctuations, and an effective interface thickness. In this abstraction, the explicit competitor class in the occupancy and charge model represents charged cfDNA-like background fragments rather than the full biochemical background of whole blood. Protein adsorption, fouling-layer growth, and other non-DNA interferents motivate the whole-blood setting. In the model, their effects enter indirectly through effective interface thickness, the blank distribution, and the detection threshold. For the charge coupling calculation, we define the lumped interface separation
\begin{equation}
d_{\mathrm{eff}} = t_{\mathrm{ox}} + d_b ,
\label{eqdeff}
\end{equation}
where $t_{\mathrm{ox}}$ is the oxide thickness and $d_b$ is the total biofunctional-layer thickness above the oxide, including the linker or silanization layer, the antifouling layer, and the receptor layer.  

The sensing area determines the maximum number of bound molecules that can contribute to the signal. Let $W$ and $L$ denote the channel width and channel length, respectively. Under a planar-area approximation, the active sensing area is $A = WL$. In the present implementation, we use the same area for electrostatic coupling. With receptor surface density $\rho_R$, the number of available binding sites is
\begin{equation}
N_R = \rho_R A.
\label{eqNR}
\end{equation}
Finite $N_R$ imposes a hard capacity limit at the sensing interface and prevents unrealistically large current shifts at higher target concentrations \cite{Movilli2018Control,Graphene_HybridizationKinetics}.

\hl{Biomolecule interaction with the BioFET surface is represented through a stochastic site assignment process with finite receptor capacity. Let $N_T$ and $N_C$ denote the numbers of target and background fragments in a fixed volume of whole blood denoted by $V_{\mathrm{sample}}$.} Let $C_t$ and $C_c$ denote their respective concentrations, where $C_c$ describes the charged cfDNA background fragments included in the surface interaction model. The expected molecular populations are calculated as $\bar{N}_T=C_tV_{\mathrm{sample}}N_A$ and $\bar{N}_C=C_cV_{\mathrm{sample}}N_A$, where $N_A$ is Avogadro's constant. For each Monte Carlo realization, the molecular populations are drawn as $N_T\sim\mathrm{Poisson}(\bar{N}_T)$ and $N_C\sim\mathrm{Poisson}(\bar{N}_C)$. In this study, $V_{\mathrm{sample}}=\SI{10}{\milli\litre}$ defines the whole blood volume used to convert the specified concentrations into molecular populations for stochastic surface assignment. Flow, circulation time, and reaction kinetics are reserved for future model extensions.

Each molecule is assigned a dimensionless binding weight. Target weights are sampled from $u_t\sim\mathcal{U}(0,1)$, and background weights are sampled from $u_c\sim\mathcal{U}(0,0.5)$. \hl{These weights encode receptor selectivity by assigning a higher binding probability to target molecules than to background fragments} \cite{Frutiger2021NSB}. \hl{The uniform distributions capture randomness within the selected interaction ranges.} Target and background fragment lengths are drawn from fixed ranges, with $N_{bp}^{(t)}\in[50,250]$ \cite{Wen2022ctDNA} and $N_{bp}^{(c)}\in[180,360]$ \cite{Shi2020cfDNASize}. These fragment ranges are selected for comparison across the modeled regimes.

Surface occupation then proceeds under the finite receptor capacity. As occupancy increases, the probability of an additional binding event is scaled by the remaining free site fraction. Target and background molecules are randomly ordered and examined sequentially. \hl{For molecule $i$ with binding weight $u_i$, the binding probability is $u_i(1-K/N_R)$, where $K$ is the current number of occupied sites. The resulting bound counts $K_t$ and $K_c$ enter the charge transduction stage. Under this interpretation, the model compares relative operating regimes under the selected binding assumptions.}

\subsection{Debye-Screened Charge-to-Current Transduction}

Electrolyte screening and interface coupling determine how strongly bound charge influences the channel. The Debye length $\lambda_D$ characterizes ionic screening from the oxide--electrolyte interface into the solution. Under physiological ionic strength, $\lambda_D$ is on the order of a nanometer, so charges extending several nanometers into the electrolyte are strongly attenuated \cite{Kesler2020DebyeBeyond,Chen2019DebyeScreening}. To compare the specified interface configurations, we use the following reduced order attenuation rule
\begin{equation}
\alpha(\lambda_D, d_{\mathrm{eff}}) = \exp\!\left(-\frac{d_{\mathrm{eff}}}{\lambda_D}\right).
\label{eqalpha}
\end{equation}
In this reduced-order model, the expression combines electrolyte screening and interface geometry into a single charge-coupling factor.

Each bound molecule contributes a surface-potential shift determined by its charge content and fragment length. Let $N_{bp,n}$ denote the number of base pairs in the $n$-th bound molecule and let $z_n$ denote an effective class-dependent charge multiplier that scales the nominal per-base-pair charge in the lumped model. In the present implementation, $z_n = z_t$ for target fragments and $z_n = z_c$ for cfDNA-like background fragments, with \hl{$(z_t,z_c)=(1,0.5)$. These selected model values assign background fragments half the effective charge contribution of target fragments. They are not derived from whole-blood measurements, and combined screening and interface attenuation are represented by $\alpha(\lambda_D,d_{\mathrm{eff}})$.}
The planar model uses a lumped interfacial capacitance density $C_{\mathrm{eff}}$ to convert bound charge into surface-potential shift. The oxide capacitance density is $C_{\mathrm{ox}}$, and the double-layer capacitance density is $C_{\mathrm{dl}}$. With effective electrolyte permittivity $\varepsilon$ and elementary charge $q$, we write
\begin{equation}
C_{\mathrm{eff}} =
\left(C_{\mathrm{ox}}^{-1} + C_{\mathrm{dl}}^{-1}\right)^{-1},
\qquad
C_{\mathrm{dl}} = \frac{\varepsilon}{\lambda_D},
\label{capacitances}
\end{equation}
so the oxide and double-layer terms enter through one lumped areal capacitance. In this lumped charge-coupling model, \hl{accounting for the nominal bare charge of $2q$ per base pair} \cite{Hedley2024}, the unscreened surface-potential shift is
\begin{equation}
\Delta \psi_n =
\frac{z_n 2q N_{bp,n}}{A\,C_{\mathrm{eff}}}.
\label{eqpotential}
\end{equation}
Eq.~\eqref{eqpotential} treats the charge of each bound DNA fragment at the selected lumped interface separation, providing a compact coupling representation for regime comparison \cite{Josephs2012ElectricField,DeVico2011Predicting}.
The screened current shift is then
\begin{equation}
\Delta I_D = g_m \sum_{n=1}^{K_t+K_c} \Delta \psi_n \,\alpha(\lambda_D, d_{\mathrm{eff}}).
\label{eq:screened_current_shift}
\end{equation}
Because detection uses current-shift magnitude, we report the noise-free signal amplitude as $|\Delta I_D|$. The measured shift used in detection is $\Delta I_D^{\mathrm{meas}} = \Delta I_D + \eta_I$.
This completes the measurement model.

\subsection{Blank-derived Threshold Detection}\label{sec_blank_thres}

We formulate detection as a binary hypothesis test on the magnitude of the sensing-induced current shift. In the absence of target molecules, the noise-free current shift arises from background fragments. The practical question at ultra-low target concentration is whether the distribution of $|\Delta I_D^{\mathrm{meas}}|$ with target present can be distinguished from the corresponding blank distribution.

The intrinsic electrical floor is attributed to thermal noise and low-frequency $1/f$ noise, which dominate BioFET operation in the present low-frequency regime \cite{Deen2006NoiseBioFET,Mori2022SNRBioFET}. The measured current noise is modeled as a zero-mean Gaussian term, $\eta_I \sim \mathcal{N}(0,I_{n,\mathrm{rms}}^2)$, where $I_{n,\mathrm{rms}}$ is obtained by integrating the total current-noise PSD over the measurement band $[f_{\min},f_{\max}]$.

Let $N_0$ denote the number of blank realizations used to estimate
the threshold. \hl{To isolate the target-induced signal from background competitor noise and device fluctuations, the detection threshold is established using a baseline derived from target-free realizations.} The total sensing-induced current shift is $\Delta I_D =
\Delta I_t +\Delta I_c$, where $\Delta I_t$ and $\Delta I_c$ are the target-induced and background-fragment-induced current shifts. In blank realizations, $\Delta I_t = 0$, so the measured blank shift is $\Delta I_{\mathrm{blank}}^{\mathrm{meas}}=\Delta I_c+\eta_I$. The detection
threshold $\theta$ is set using a Gaussian approximation to the upper one-sided 95th percentile of the blank-shift magnitudes. In the implementation, $\theta$ is computed as the mean blank magnitude plus 1.645 times its standard deviation \cite{Mohamad2018LOBLODLOQ}. For each simulated operating regime, this threshold is estimated from the corresponding blank realizations and then held fixed for target-present trials in that regime.

Let $M$ denote the number of sensors. A sensor-level detection
event is declared when
\begin{equation}
\delta_{i,m} = \mathbb{1}\!\left\{ \left| \Delta I_{D,i,m}^{\mathrm{meas}} \right| > \theta \right\},
\label{eqDetection}
\end{equation}
where $\delta_{i,m}\in\{0,1\}$ is the binary decision of sensor $m$ in Monte Carlo realization $i$. OR fusion across the $M$ sensors gives the realization-level decision, 
\begin{equation}
\hat{y}_i = \mathbb{1}\!\left\{ \sum_{m=1}^{M} \delta_{i,m} \ge 1 \right\},
\label{eqsampledecision}
\end{equation}
where $\hat{y}_i \in \{0,1\}$ denotes the final binary decision for realization $i$. \hl{For $M=2$, a logical OR rule declares a positive result when either sensor crosses the threshold. Sensitivity and specificity are then computed from target-present and blank realizations, respectively. Extension to larger $M$ would require separate evaluation of the threshold and fusion rule.}

With the realization-level decisions defined, sensitivity and specificity follow directly from the target-present and blank realizations. Sensitivity is the fraction of target-present realizations with $\hat{y}_i=1$. Specificity is one minus the fraction of blank realizations with $\hat{y}_i=1$, where $\hat{y}_i=1$ is a false positive in the blank condition. With $N_1$ target-present realizations and $N_0$ blank realizations, we compute
\begin{equation}
\mathrm{Sensitivity} = 100 \cdot \frac{1}{N_1}\sum_{i=1}^{N_1}\hat{y}_i,
\qquad
\mathrm{Specificity} = 100 \cdot \left(1 - \frac{1}{N_0}\sum_{i=1}^{N_0}\hat{y}_i\right).
\label{eqmetrics}
\end{equation}
These metrics are used throughout Section~\ref{sec_SimResults} to summarize detection performance.

\section{Simulation Results and Analysis}\label{sec_SimResults}

All simulations were implemented in MATLAB R2025b using the model defined in Section~\ref{sec_Framework}. Table~\ref{tab_sim_params} lists the parameters used in the reported simulations. Unless stated otherwise, in each sweep, the swept variable is varied while the remaining parameters are fixed to the values in Table~\ref{tab_sim_params}. \hl{The values for $t_{ox}$ and $d_b$ are chosen to reflect typical BioFET design parameters} \cite{Layasree2021SimulationAC,Gao2015HighIonicStrength,Kesler2020DebyeBeyond}. The binding weights are selected model parameters that encode relative target and background interaction strengths. Accordingly, the results characterize operating regimes under the stated assumptions.

\begin{table}[!t]
\centering
\caption{Simulation parameters.}
\vspace{-1em}
\label{tab_sim_params}
\small
\setlength{\tabcolsep}{4pt}
\renewcommand{\arraystretch}{1.15}
\begin{tabular}{p{0.64\columnwidth} p{0.28\columnwidth}}
\hline
\textbf{Parameter} & \textbf{Value} \\
\hline
Operating drain bias, $V_{SD}$ &  \SI{0.1}{\volt} \\
Operating gate bias, $V_{SG}$ & \SI{0.3}{\volt} \\
Transconductance, $g_m$ & \SI{1.242e-6}{\siemens} \\
Baseline drain current, $I_{D,0}$ & \SI{4.306e-08}{\ampere}  \\
Channel width, $W$ & \SI{670}{\nano\meter} \\
Channel length, $L$ & \SI{1}{\micro\meter} \\
Oxide thickness, $t_{\mathrm{ox}}$ & \qtyrange{2}{5}{\nano\meter} \\

Biofunctional-layer thickness, $d_b$ & \qtyrange{5}{9}{\nano\meter} \cite{Gao2015HighIonicStrength,Kesler2020DebyeBeyond}\\
Receptor surface density, $\rho_R$ & $\sim$ \qty{1e12}{\per\centi\metre\squared} \cite{Movilli2018Control}\\
Target concentration, $C_t$ & \qty{0.1}{\atto\molar} to \qty{1}{\femto\molar} \cite{Wen2022ctDNA}\\
Background fragment concentration, $C_c$ & \SI{1}{\femto\molar} \cite{Shi2020cfDNASize}\\
Effective charge multipliers, $z_t, z_c$ & $(1, 0.5)$ \\

Debye length, $\lambda_D$ & \qtyrange{0.7}{1.5}{\nano\meter} \cite{Kesler2020DebyeBeyond}\\
Fixed whole blood volume, $V_{\mathrm{sample}}$ & \SI{10}{\milli\litre} \\

Bandwidth, $[f_{\min},f_{\max}]$ & (\SI{1}{\hertz}, \SI{1000}{\hertz}) \\

Number of sensors, $M$ & 2 \\
Number of Monte Carlo, $N$ & $1000 (= N_0 =N_1)$ \\
\hline 
\end{tabular} \vspace{-1em}
\end{table}

\subsection{Screening-limited Signal Analysis}

\begin{figure}[!t]
    \centering
    \includegraphics[width=\linewidth]{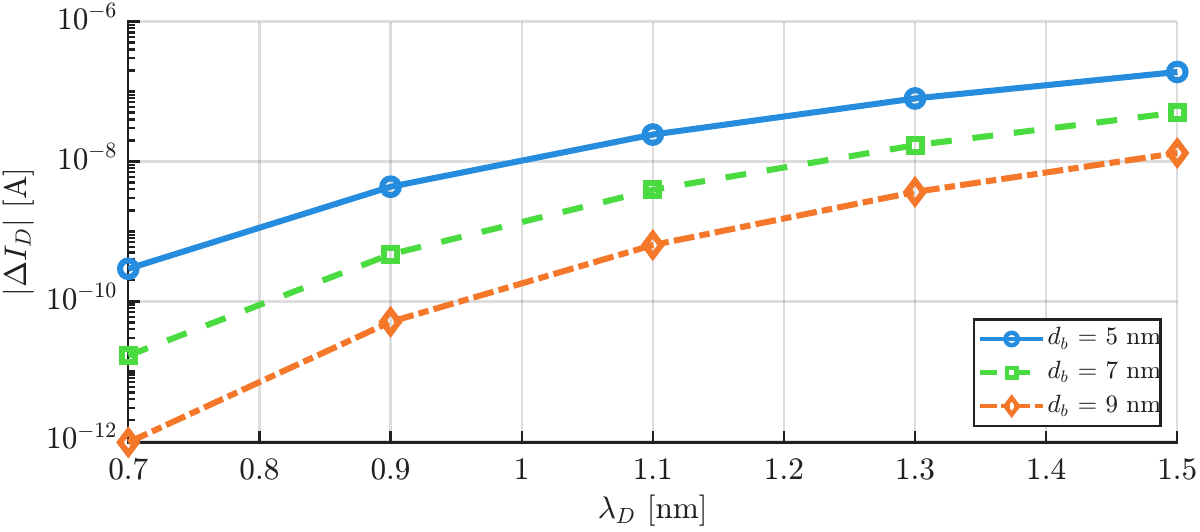}
    \caption{Signal amplitude $|\Delta I_D|$ versus $\lambda_D$ for varying values of $d_b$ with $t_{\mathrm{ox}}=3.5$~nm, $C_t=0.1$~aM, and $M=1$.}    \label{fig:signal_vs_debye}
\end{figure}

Fig.~\ref{fig:signal_vs_debye} isolates screening-limited signal formation for $M=1$ at $C_t=0.1$~aM. For all three interface thicknesses, $|\Delta I_D|$ increases with $\lambda_D$ because a longer Debye length increases the modeled charge coupling. At $\lambda_D=0.7$~nm, increasing $d_b$ from 5~nm to 7~nm and 9~nm lowers the current shift by roughly one order of magnitude and more than two orders of magnitude, respectively. The separation narrows as $\lambda_D$ increases, while $d_b=5$~nm retains the largest signal throughout the sweep. These results show that a shorter biofunctional layer and a longer Debye length provide stronger modeled coupling before thresholding and electronic noise are applied. This trend agrees with prior observations that high ionic strength and long charge-to-channel separation strongly weaken charge-based BioFET sensing \cite{Gao2015HighIonicStrength,Stern2010WholeBlood,Kesler2020DebyeBeyond}..

\begin{figure}[!t]
    \centering
    \includegraphics[width=0.48\textwidth]{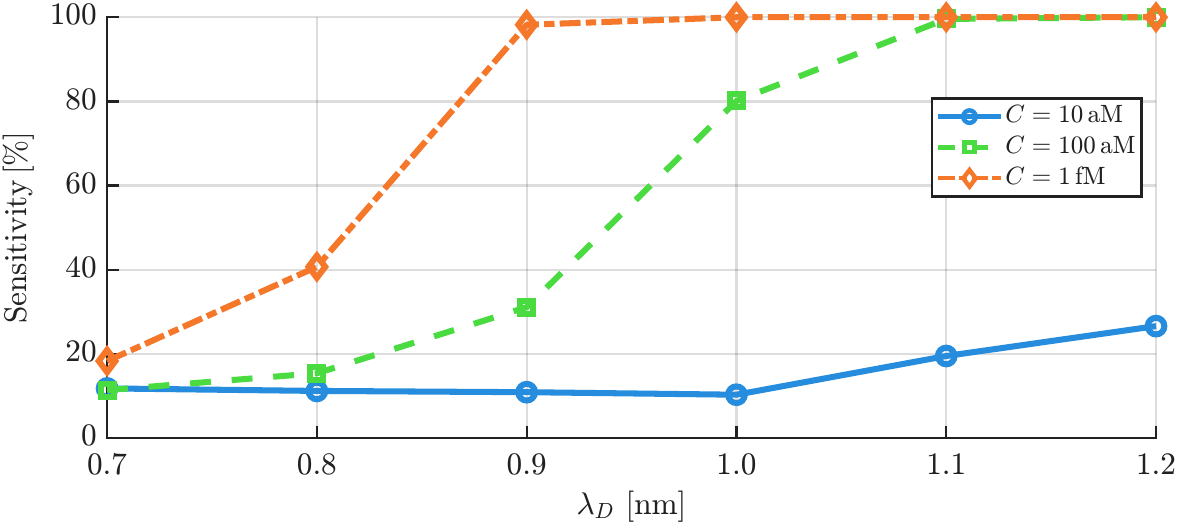}
    \caption{Sensitivity versus $\lambda_D$ for three target concentrations with $t_{\mathrm{ox}} = 3.5$~nm and $d_b = 5$~nm.} 
    \label{fig:sens_vs_debye}
\end{figure}

Fig.~\ref{fig:sens_vs_debye} translates the signal trend into detection performance. At 1~fM, sensitivity increases from approximately 18\% at $\lambda_D=0.7$~nm to 40\% at 0.8~nm, 98\% at 0.9~nm, and reaching 100\% at 1~nm. At 100~aM, it rises from about 10\% at 0.7~nm to 31\% at 0.9~nm and 80\% at 1.0~nm, reaching 100\% at 1.1~nm. At 10~aM, it increases more gradually, from approximately 10\% to 25\% across the same range. These transitions depend strongly on target concentration and show that longer Debye length expands the operating regime.

\begin{figure}[!t]
    \begin{subfigure}[t]{\linewidth}
        \centering
        \includegraphics[width=\linewidth]{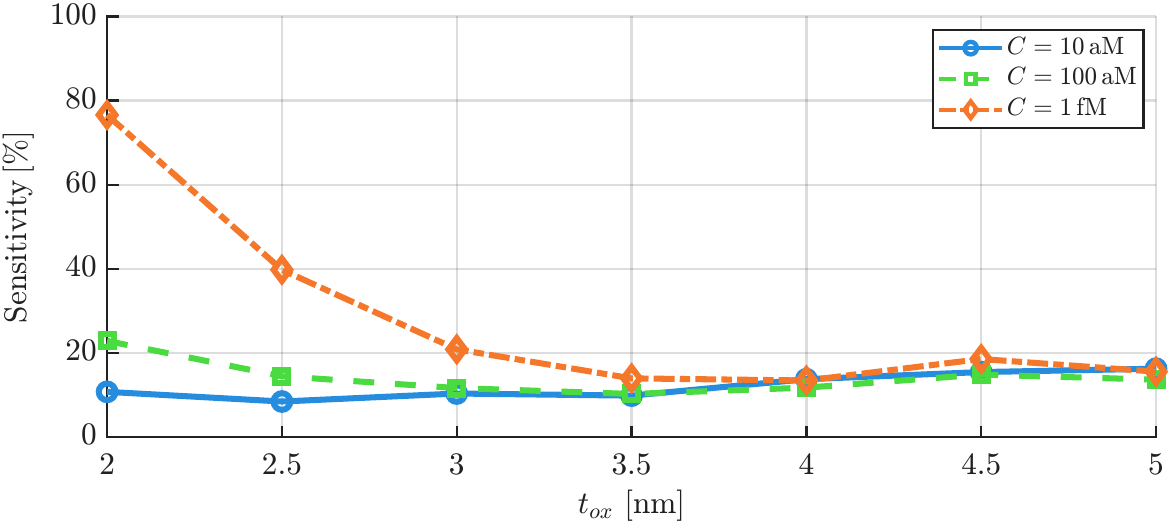} \vspace{-1em}
        \label{fig:sensitivity5db}
        \caption{}
    \end{subfigure}
    \begin{subfigure}[t]{\linewidth}
            \centering
            \includegraphics[width=\linewidth]{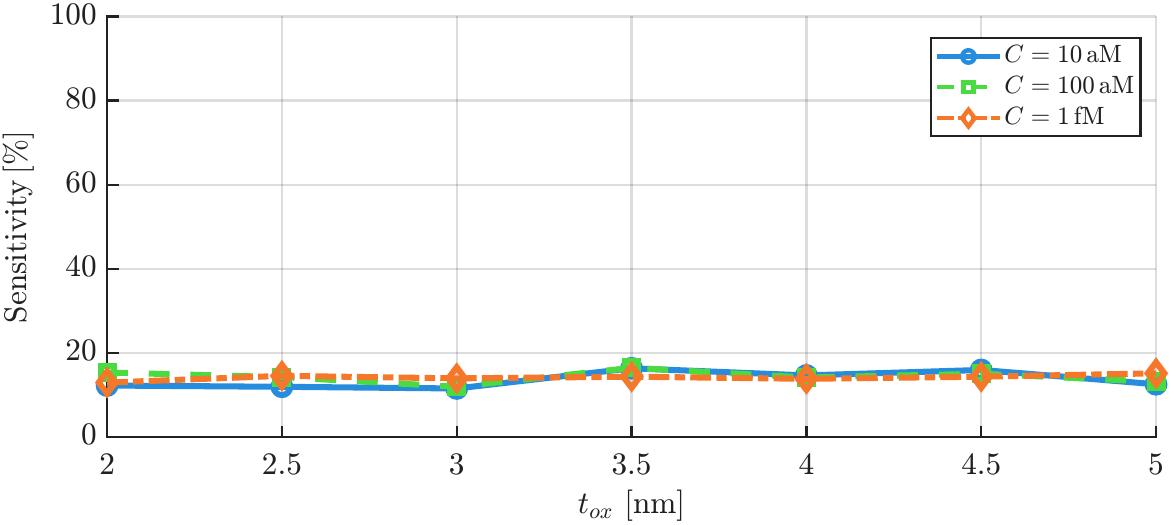}
            \caption{}
            \label{fig:sensitivity7db}
    \end{subfigure}    
    \caption{Sensitivity versus $t_{\mathrm{ox}}$ for varying target concentrations with $\lambda_D = 0.7$~nm and (a) $d_b = 5$~nm, (b) $d_b = 7$~nm.} \vspace{-1em}
    \label{fig:full-sens-analysis}
\end{figure}

\subsection{Sensitivity and Specificity Analysis}

Fig.~\ref{fig:full-sens-analysis} shows the modeled sensitivity to oxide thickness under strong screening with $\lambda_D=0.7$~nm. For $d_b=5$~nm, the 1~fM sensitivity decreases from approximately 76\% at $t_{\mathrm{ox}}=2$~nm to 40\% at 2.5~nm and remains within approximately 14\%--20\% from 3~nm to 5~nm. The 100~aM and 10~aM cases remain within approximately 8\%--22\% across the sweep. For $d_b=7$~nm, all three concentration curves cluster around 10\%--16\% and show no sustained trend with oxide thickness. This comparison indicates that a shorter biofunctional layer provides the largest sensitivity gain for the 1~fM case with a thin oxide.

\begin{figure}[!t]
    \centering
            \includegraphics[width=\linewidth]{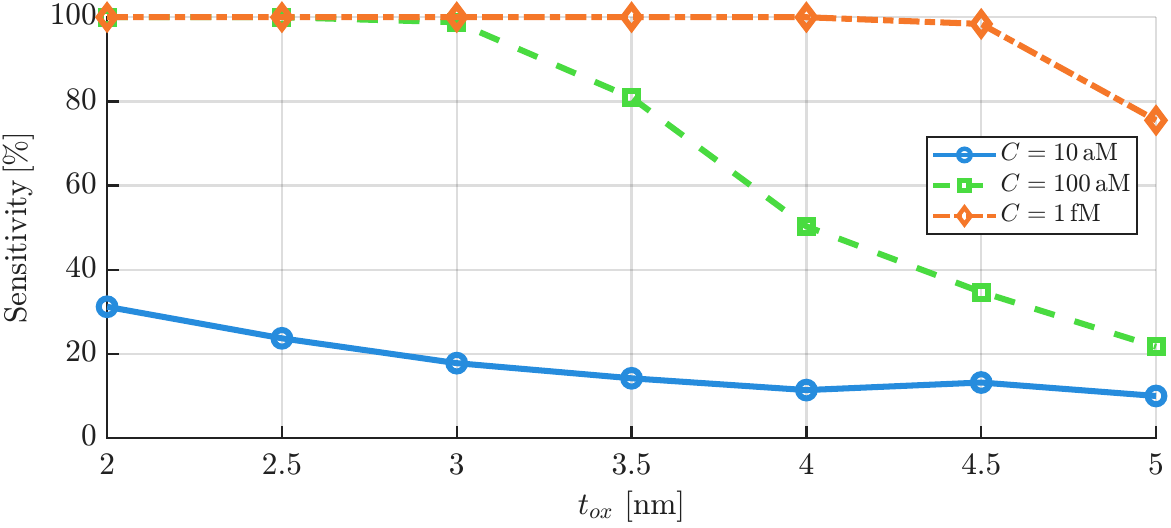}  
        \caption{Sensitivity versus $t_{\mathrm{ox}}$ for varying target concentrations with $d_b = 5$~nm and $\lambda_D = 1$~nm.} \vspace{-1em}
        \label{fig:sens-oxide-debye}
\end{figure}

Fig.~\ref{fig:sens-oxide-debye} repeats the oxide-thickness sweep at $\lambda_D=1$~nm. At 1~fM, sensitivity remains near 100\% from $t_{\mathrm{ox}}=2$~nm to 4.5~nm and decreases to approximately 76\% at 5~nm. At 100~aM, sensitivity remains near 100\% from 2~nm to 3~nm, followed by a decrease to approximately 81\%, 50\%, 34\%, and 22\% at oxide thicknesses of 3.5, 4, 4.5, and 5~nm, respectively. At 10~aM, sensitivity decreases from approximately 31\% at 2~nm to 10\% at 5~nm, with small stochastic fluctuations. Comparison with Fig.~\ref{fig:full-sens-analysis}(a) shows that the longer Debye length expands the range of conditions that provide high sensitivity. The results also show continued oxide-thickness dependence at the lower target concentrations.

\begin{figure}[!t]
    \centering
            \includegraphics[width=\linewidth]{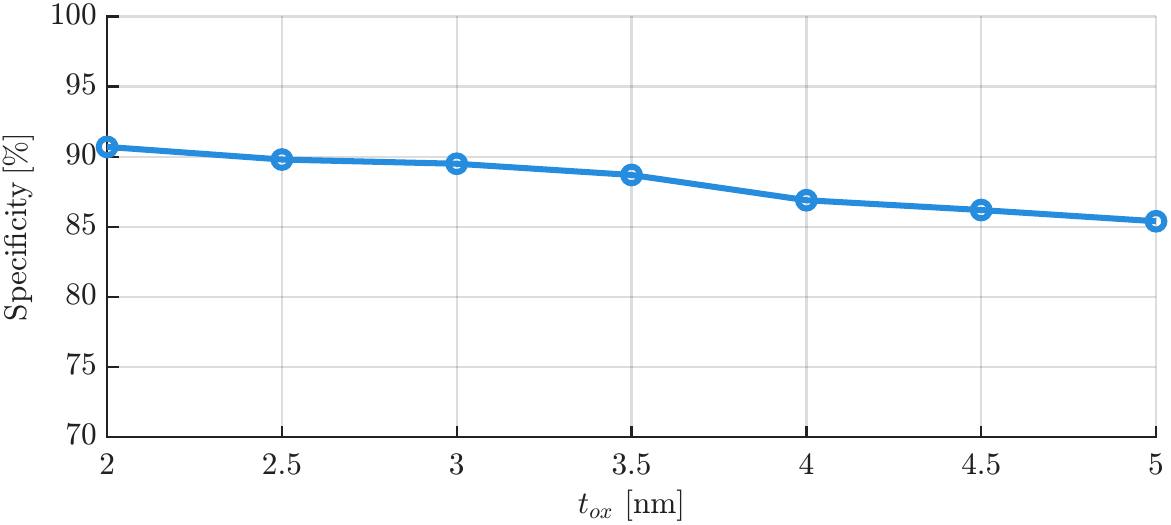}  
        \caption{Specificity versus $t_{\mathrm{ox}}$ with $d_b = 5$~nm and $\lambda_D = 0.7$~nm.}
        \label{fig:specificity}
\end{figure}

Fig.~\ref{fig:specificity} shows a gradual decrease in specificity from approximately 90.5\% at $t_{\mathrm{ox}}=2$~nm to 85.5\% at $t_{\mathrm{ox}}=5$~nm. The corresponding blank false-positive fraction increases from approximately 9.5\% to 14.5\% under the selected threshold and $M=2$ OR rule. Across the same strong-screening geometry range (\textit{cf.} Fig.~\ref{fig:full-sens-analysis}), sensitivity spans approximately 10\%--76\%, depending on target concentration and biofunctional-layer thickness. Together, these metrics characterize the balance between target detection and blank rejection in the exploratory model. Section~\ref{sec_noise} next examines the electronic noise contribution.

\begin{figure}[!t]
    \centering
        \includegraphics[width=\linewidth]{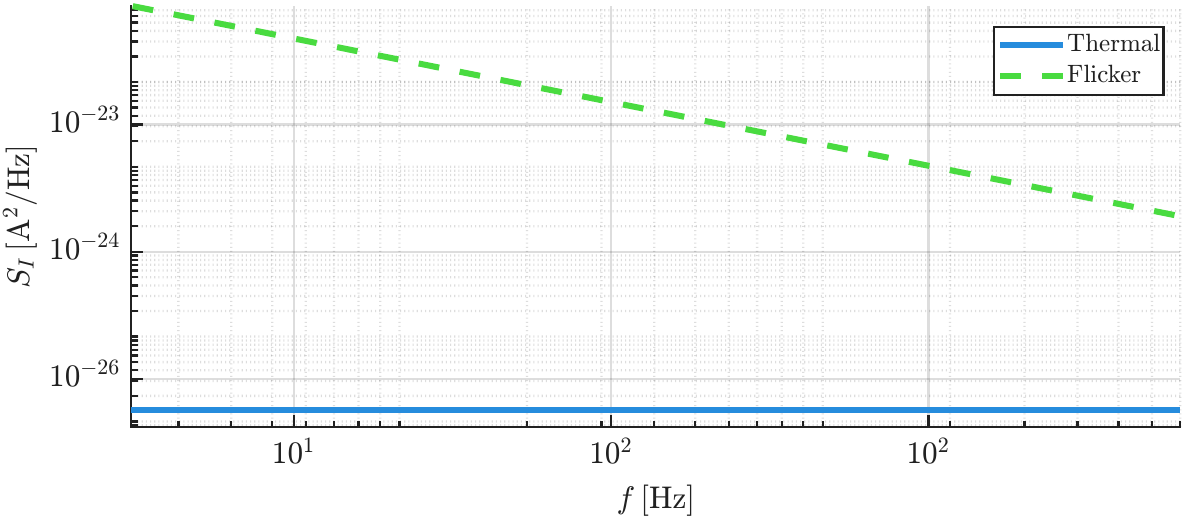}
        \caption{Thermal and $1/f$ current noise power spectral density for $t_{\mathrm{ox}} = 3.5$~nm.} \vspace{-1em}
    \label{fig:noise_analysis}
\end{figure}

\subsection{Low-frequency Noise Analysis}\label{sec_noise}

Fig.~\ref{fig:noise_analysis} shows the electronic noise floor used in the detection model. The thermal component remains approximately constant at the order of $10^{-27}$~A$^2$/Hz. The flicker component decreases from the order of $10^{-23}$~A$^2$/Hz at 1~Hz to the order of $10^{-25}$~A$^2$/Hz at 1~kHz and remains at least two orders of magnitude above the thermal component throughout the measurement band. Flicker noise therefore provides the dominant electronic noise contribution in the modeled quasi-static operating regime. When screened current shifts approach the blank distribution, the integrated noise further reduces the separation between target-present and blank responses.

\section{Conclusions and Future Work}\label{sec_conclusion}

This exploratory study identifies interface and screening conditions that produce distinct ctDNA detection regimes in the modeled whole-blood environment. Longer Debye length and shorter interface separation expand the range of conditions that provide high sensitivity, particularly at 100~aM and 1~fM. Stronger screening and larger interface separation produce lower sensitivity across much of the tested range. Across the evaluated oxide-thickness sweep at $\lambda_D=0.7$~nm and $d_b=5$~nm, specificity remains within approximately 85\%--91\% under the selected blank threshold and two-sensor OR rule. These results frame BioFET-based ctDNA sensing as a constrained operating regime shaped by screening, interface geometry, target concentration, background fluctuations, and low-frequency noise.

The model focuses on boundary behavior through a quasi static local sensing description, selected binding weight ranges, a fixed whole blood volume, and a blank threshold detector. Future extensions can add time dependent transport and receptor kinetics, interface evolution, broader noise mechanisms, fabrication uncertainty, and threshold sweeps. The attenuation factor applies the electrolyte Debye length to the lumped quantity $d_{\mathrm{eff}}=t_{\mathrm{ox}}+d_b$. Under this coupling rule, the oxide thickness trends provide a reduced order view of the combined influence of electrolyte screening and interface geometry. The reported sensitivity curves characterize operating regimes under the specified volume and receptor selectivity assumptions.

Future work should test whether thinner interface stacks, improved surface chemistries, alternative device structures, or different threshold choices can widen the separation between target-present and blank responses.
More broadly, closing the ctDNA whole-blood gap may require interface improvements within the BioFET architecture and closer examination of transduction principles that avoid electrostatic field gating in physiological electrolytes. For intrabody nanonetworks, the key takeaway is that local sensing performance must be evaluated in whole blood. In the BioFET case studied here, physiological screening, interface distance, background binding, and noise can reduce the target signal enough to limit reliable ctDNA detection.

\begin{acks}
This project has received funding from the European Union's Horizon Europe research and innovation programme under the Marie Sk\l odowska-Curie grant agreement No.~101154851.
\end{acks}

\printbibliography
\end{document}